\documentclass[12pt,preprint]{aastex}
\usepackage{mkfig}
\begin{document}

\title{\bf Resolving the $\xi$~Boo Binary with {\em Chandra}, and Revealing
  the Spectral Type Dependence of the Coronal ``FIP Effect''}

\author{Brian E. Wood\altaffilmark{1}, Jeffrey L. Linsky\altaffilmark{2}}


\altaffiltext{1}{Naval Research Laboratory, Space Science Division,
  Washington, DC 20375; brian.wood@nrl.navy.mil}
\altaffiltext{2}{JILA, University of Colorado and NIST, Boulder, CO
  80309-0440; jlinsky@jila.colorado.edu}

\begin{abstract}

     On 2008 May 2, {\em Chandra} observed the X-ray spectrum of
$\xi$~Boo (G8~V+K4~V), resolving the binary for the first time
in X-rays and allowing the coronae of the two stars to be studied
separately.  With the contributions of $\xi$~Boo~A and B to the
system's total X-ray emission now observationally established (88.5\%
and 11.5\%, respectively), consideration of mass loss measurements for
GK dwarfs of various activity levels (including one for $\xi$~Boo)
leads to the surprising conclusion that $\xi$~Boo~B may dominate the
wind from the binary, with $\xi$~Boo~A's wind being very weak despite
its active corona.  Emission measure distributions and coronal
abundances are computed for both stars and compared with {\em Chandra}
measurements of other moderately active stars with G8-K5 spectral
types, all of which exhibit a narrow peak in emission measure near
$\log T=6.6$, indicating that the coronal heating process in these
stars has a strong preference for this
temperature.  As is the case for the Sun and many other stars, our
sample of stars shows coronal abundance anomalies dependent on the
first ionization potential (FIP) of the element.  We see no dependence
of the degree of ``FIP effect'' on activity, but there is a dependence
on spectral type, a correlation that becomes more convincing when
moderately active main sequence stars with a broader range of spectral
types are considered.  This clear dependence of coronal abundances on
spectral type weakens if the stellar sample is allowed to be
contaminated by evolved stars, interacting binaries, or extremely
active stars with $\log L_X>29$, explaining why this correlation has
not been recognized in the past.

\end{abstract}

\keywords{stars: individual ($\xi$~Boo) --- stars: coronae --- stars:
  late-type --- X-rays: stars}

\section{INTRODUCTION}

          At a distance of only 6.7~pc, the primary of the $\xi$~Boo
binary system, $\xi$~Boo~A (G8~V), is one of the two brightest
``moderately active'' coronal X-ray sources in the sky.  The only star
of this coronal activity level with a comparable X-ray flux at Earth
is $\epsilon$~Eri (K2~V, $d=3.2$~pc).  As such, these two stars are of
crucial importance for establishing the coronal properties of stars
that are significantly more active than the Sun, but which are not in
the extremely active regime that would be represented by the RS~CVn
binaries or very young stars with rotation periods of only a few days.

     \citet{jjd00} find that the coronal emission measure (EM)
distributions of $\epsilon$~Eri and $\xi$~Boo are similar to those of
bright active regions on the Sun.  They note that the X-ray emission
from these stars can therefore be explained by stellar surfaces that
are completely covered with solar-like active regions, suggesting that
$\epsilon$~Eri and $\xi$~Boo~A represent a maximum of solar-like
activity for coronal stars, whereas stars with even higher X-ray
luminosities and higher coronal temperatures represent a completely
different regime of coronal activity.  This interpretation places
$\epsilon$~Eri and $\xi$~Boo~A at an interesting coronal transition
point.

     Further evidence that $\epsilon$~Eri and $\xi$~Boo~A lie near a
transition in the nature of coronal activity comes from stellar wind
measurements, which are based on analyses of H~I Ly$\alpha$ absorption
from wind-ISM interaction regions \citep{bew05a}.  At low
activity levels mass loss increases with activity, with $\epsilon$~Eri
having a mass loss rate 30 times that of the Sun
(i.e., $\dot{M}=30~\dot{M}_{\odot}$, where
$\dot{M}_{\odot}=2\times 10^{-14}$ M$_{\odot}$~yr$^{-1}$).  However,
the somewhat more active $\xi$~Boo system has a surprisingly weak
wind with only $5~\dot{M}_{\odot}$, leading to the suggestion that
the observed increase of mass loss with activity ends abruptly at
an X-ray surface flux of about $F_X=1\times 10^{6}$ ergs cm$^{-2}$
s$^{-1}$, with $\epsilon$~Eri and $\xi$~Boo on opposite sides of this
dividing line \citep{bew05a}.

     The wind measurements suggest a slight modification of the
proposal that $\epsilon$~Eri and $\xi$~Boo~A represent an extreme of
truly solar-like activity.  They suggest that $\epsilon$~Eri is the
true extreme, while $\xi$~Boo~A actually lies beyond it at the low end
of the non-solar-like activity regime.  Evidence that
$\xi$~Boo~A's magnetic topology is fundamentally different from the
Sun comes from spectropolarimetric measurements of photospheric fields
on $\xi$~Boo~A.  \citet{pp05} find evidence for both a global
dipole field component much stronger than that of the Sun ($\sim
40$~G), and a strong toroidal field component ($\sim 120$~G) that has
no clear solar analog.  We also note that the activity level of
$\xi$~Boo~A is roughly where one starts to see polar starspots
\citep{kgs02}, which also have no solar analog.  \citet{cgt88} find
evidence for high latitude spots on $\xi$~Boo~A, though not
necessarily ``polar spots.''  Perhaps the strong large-scale fields
found by \citet{pp05} envelope the star and inhibit wind flow,
explaining why $\xi$~Boo's wind is surprisingly weak.  Consistent with
this idea, \citet{pp08} present evidence that as stellar
rotation and magnetic activity increase, stars tend to store a larger
fraction of their magnetic energy in large scale fields, which could
in principle inhibit mass loss.

     With this background in mind, we here present an analysis of the
X-ray spectrum of $\xi$~Boo~A based on new observations from {\em
Chandra}, with the goal of seeing whether the apparent changes in
coronal character noted above find clear expression in the coronal
spectrum of $\xi$~Boo~A, and coronal properties inferred from it.  A
similar spectrum of $\epsilon$~Eri has already been
analyzed, so a central goal of this paper is to compare the
$\xi$~Boo~A spectrum with that of $\epsilon$~Eri and similar stars
\citep[][hereafter WL06]{bew06}.

     Other coronal spectra of $\xi$~Boo have been obtained by previous
X-ray and EUV missions \citep[e.g.,][]{jml99,jcp09},
but the new {\em Chandra} data are superior in many
ways.  One of them is that {\em Chandra}/LETGS observations
allow the X-ray spectrum to be observed at high
spectral resolution over a very broad spectral range of $5-175$~\AA.
This allows the detection of many coronal emission lines that
have never been previously detected before from $\xi$~Boo~A.

    Another particulary important improvement is that {\em Chandra}'s
superb spatial resolution allows us for the first
time to truly address the issue of $\xi$~Boo's binarity.  For
$\xi$~Boo~A has a companion, $\xi$~Boo~B (K4~V), which is only
6$^{\prime\prime}$ away, and {\em no X-ray or UV observation of
$\xi$~Boo has ever truly resolved the binary.}  Table~1 lists various
properties of the two $\xi$~Boo stars.  The faster rotation and larger
size of the primary lead to the expectation that $\xi$~Boo~A should
dominate the coronal emission from the system.  Detailed analyses of
IUE data and a ROSAT/HRI image of $\xi$~Boo do indeed suggest that
$\xi$~Boo~A dominates the binary's UV and X-ray emission
\citep{lh79,js97}, so the usual assumption of
simply assigning all flux to the primary should be a decent
approximation.  However, $\xi$~Boo~B's spectral type and rotation
period are not that different from the X-ray luminous $\epsilon$~Eri
(K2~V; $P_{rot}=11.7$~days).  Thus, $\xi$~Boo~B should be contributing
a non-negligible flux to the system's total X-ray emission.  The older
observations simply do not have sufficient spatial resolution to
quantify how much.  In short, the new {\em Chandra} data presented
here allow the first X-ray detection of $\xi$~Boo~B, and they ensure that
the brighter spectrum of $\xi$~Boo~A is uncontaminated by any emission
from the companion.

\section{OBSERVATIONS AND DATA REDUCTION}

     The $\xi$~Boo binary was observed on 2007~May~2 for 97.2~ksec
with {\em Chandra}'s LETGS configuration, combining the Low Energy
Transmission Grating (LETG) and the component of the High Resolution
Camera detector intended for use in spectroscopy (HRC-S).  The
zeroth-order LETGS image of $\xi$~Boo is displayed in Figure~1,
showing how cleanly the stars are separated in the {\em Chandra} data,
representing the first resolution of the binary in X-rays.  We measure
a position angle and stellar separation for the binary of
$\theta=310.9^{\circ}$ and $\rho=6.30^{\prime\prime}$, respectively, in
reasonable agreement with the $\theta=310.1^{\circ}$ and
$\rho=6.15^{\prime\prime}$ values expected from the system's orbital
elements \citep{ss99}.  The zeroth order image provides the simplest
way to measure the contribution of each star to $\xi$~Boo's total
X-ray emission.  The primary dominates the emission from the system,
accounting for 88.5\% of the total counts.

     We also use the zeroth order image to assess the variability of the
two stars during the course of the daylong observation.  Figure~2 shows
light curves for the two $\xi$~Boo components, indicating significant
variability for both stars.  For $\xi$~Boo~A, there appears to be a
long duration flare near the beginning of the observation, and a
much shorter duration event later on.  In extracting spectra from the
raw {\em Chandra} data, we initially tried to separate flaring and
quiescent times.  However, we were unable to discern any clear differences
between the resulting noisy flare spectrum and that of quiescence.  Thus,
in the final spectral extraction procedure described below, we simply
extract a single spectrum for each star, including both quiescent and
flaring times.

     The standard pipeline processing of LETGS data does not
properly separate the spectra of the two stars, so the spectral
extraction must be performed in a more manual fashion.  In a previous
paper we have described in detail how we extracted separate LETGS spectra
of stars in two other binary systems, 36~Oph and 70~Oph (WL06).
We follow a very similar procedure here, which is now described in
abbreviated form.

     The data are processed using version 4.1.2 of the CIAO software.
The processing includes a removal of background counts using the
standard ``light'' pulse height background filter available in CIAO,
but we also account for time-dependent gain corrections using software
outside of CIAO, corrections which would later be added to a subsequent
official CIAO release (version 4.2).  By reducing
background counts from our data, this filtering significantly
improves the signal-to-noise of our final spectra.

     The processing initially yields a zeroth order image of the
target at the aimpoint, which we have already discussed (see Fig.~1),
and two essentially identical spectra dispersed in opposite directions
along the long axis of the HRC-S detector, a plus order and a minus
order spectrum.  The spatial resolution in the cross-dispersion
direction worsens at long wavelengths where the spectrum is furthest
from the focal point.  We use a conservatively broad source extraction
window with a width of 25 pixels centered on each stellar spectrum for
$\lambda <90$~\AA.  The window is broadened for $\lambda >90$~\AA.  To
avoid overlapping source windows, for each star we expand the window
only in the direction away from its companion star by 30 pixels.  At
the longest wavelengths, we have to accept that there is some small
degree of unresolvable blending.  We use 90 pixel background windows
on each side of the stellar spectra to determine a wavelength dependent
background to subtract from the spectra of the two stars.  Finally,
the plus and minus order spectra are coadded to yield the final
product shown in Figure~3.

\section{IMPLICATIONS FOR $\xi$~BOO'S WIND MEASUREMENT}

     The clear separation of the $\xi$~Boo binary allows the first
precise and unambiguous measurement of the X-ray luminosity of both
stars.  The system's X-ray luminosity reported in the ROSAT all-sky
survey is $\log L_X=28.91$ \citep{js04}.  Dividing this emission
between $\xi$~Boo's two stars according to the contribution ratio
suggested by Figure~1 yields the ROSAT X-ray luminosities
reported in Table~1.

     This new knowledge of the coronal activity of both $\xi$~Boo
stars has interesting ramifications for the interpretation of $\xi$~Boo's
wind measurement.  As mentioned in \S1, $\xi$~Boo has a measured wind
strength of $5~\dot{M}_{\odot}$, but the measurement is for the
combined wind of the system, and there is no way to tell from the
observations the contribution of each star to this total
\citep{bew05a}.  Figure~4 plots mass loss rate as a function of
coronal X-ray emission for all main sequence stars with measured
winds, analogous to a figure from \citet{bew05a}.  But in this
figure we plot separate points for $\xi$~Boo~A and $\xi$~Boo~B, for
three different assumptions about how the collective wind is divided
between the two stars.

     We now know for a fact that $\xi$~Boo~B is very similar to
$\epsilon$~Eri, 36~Oph, and 70~Oph in terms of its coronal activity.
But these stars all have rather high mass loss rates.  The only way
that $\xi$~Boo~B can be consistent with even the weakest of these
winds, that of 36~Oph, is if $\xi$~Boo~B accounts for practically all
of $\xi$~Boo's wind.  This would leave the more active $\xi$~Boo~A
with a wind that is possibly weaker than that of the Sun, despite its
active corona, strengthening the case for the existence of a wind
dividing line near an X-ray surface flux of $F_X=1\times 10^{6}$ ergs
cm$^{-2}$ s$^{-1}$ (see Fig.~4), where stronger large scale magnetic
fields begin to inhibit mass loss.  Even if $\xi$~Boo~A accounts for
nearly all of the binary's wind, it would still be weak compared to
that of $\epsilon$~Eri, and then one would have to explain why
$\xi$~Boo~B apparently has a wind over an order of magnitude weaker
than that of many other stars of equal activity.

     We come to the surprising conclusion that the most likely
interpretation of the available data is that $\xi$~Boo~B dominates
the wind of the $\xi$~Boo system.  This conclusion is also consistent
with the idea that a wind dividing line exists at an X-ray surface flux
of about $F_X=1\times 10^{6}$ ergs cm$^{-2}$ s$^{-1}$ (see \S1),
separating less active stars with strong winds and more active stars
with weaker winds \citep{bew05a,bew04}.  This activity dependent
wind dividing line is not to be confused with the spectral type
dependent wind-corona dividing line long known to exist for red
giant stars \citep{jll79}.  It must be noted that our conclusions rely
on a very limited number of wind detections.  Further support for
these interpretations should be sought from more wind measurements
of main sequence stars, particularly for active stars.

\section{SPECTRAL ANALYSIS}

\subsection{Line Identification and Measurement}

     We use version 6.0 of the CHIANTI atomic database to identify
emission lines in our spectra \citep{kpd97,kpd09}.  The
identified lines are shown in Figure~3 and listed in Table~2.  Table~2
lists counts for the detected lines, measured by direct integration
from the spectra.  Line formation temperatures are quoted in the third
column of the table, based on maxima of contribution functions
computed using the ionization equilibrium computations of \citet{ma85}.
Uncertainties quoted in Table~2 are 1$\sigma$.
For the fainter $\xi$~Boo~B source, only the strongest lines are
detected, but we quote 2$\sigma$ upper limits for the
nondetections, based on summing in quadrature the uncertainties in 5
bins surrounding the location of the line, and then multiplying this
sum by two.

     Measurements are made for all lines that appear visually to be at
least marginally detected.  However, cases in which the uncertainty
implies a $<2\sigma$ detection must be considered questionable.  After
the emission measure analysis described in \S4.4 is completed, we confirm
{\em a posteriori} that these questionable detections are at least
plausible based on the derived emission measure distributions and
abundances, which are constrained primarily by the stronger emission
lines in the analysis.

     Many of the emission lines are blends.  We
list in Table~2 all lines that we believe contribute a significant
amount of flux to the feature based on the line strengths in the
CHIANTI database.  This determination is reassessed after EM
distributions are estimated, and model spectra can be computed
from them and compared with the data.  A few of the features
identified in Figure~3 are blends of lines of different species (see
Table~2).  Although we measure counts for these blends and list them
in the table, these measurements are not used in the emission measure
analysis.

\subsection{Coronal Densities}

     Coronal electron densities can be estimated for the two $\xi$~Boo
stars directly from the O~VII $\lambda$21.8 line measurements in
Table~2.  There are in fact several He-like triplets in the observed
spectral range that are useful for density measurements: Si~XIII
$\lambda$6.7, Mg~XI $\lambda$9.2, Ne~IX $\lambda$13.6, and O~VII
$\lambda$21.8.  However, insufficient spectral resolution, low
signal-to-noise, and blending issues complicate any attempt to use the
Si~XIII, Mg~XI, and Ne~IX lines for this purpose \citep{jun02}.
For this reason, we focus only on the stronger, better separated O~VII
lines here.

      The specific line ratio of interest is that of the forbidden
line to the intersystem line:  $f/i\equiv \lambda22.101/\lambda21.807$.
Using the results of collisional equilibrium models from \citet{dp01},
the measured line ratios for $\xi$~Boo~A and $\xi$~Boo~B
translate to upper limits of $\log n_e<10.24$ and $\log n_e<11.43$,
respectively, as reported in Table~3.  The $\xi$~Boo~A limit is
consistent with values that \citet{jml99} estimate from
EUVE spectra.  Densities of about $\log n_e\approx 10$ are consistent with
spectra of $\epsilon$~Eri, 36~Oph~AB, and 70~Oph~AB as well, which have
activity levels similar to that of the $\xi$~Boo stars (WL06).
Thus, there is no clear evidence for coronal density
differences among moderately active stars observed by {\em Chandra}.

\subsection{Comparing the Line Fluxes of $\xi$~Boo~A and $\epsilon$~Eri}

     Among moderately active GK dwarfs with intrinsic X-ray luminosities
of $\log L_X=28-29$, the two easiest to observe from Earth are
$\xi$~Boo~A (G8~V) and $\epsilon$~Eri (K2~V), due to their close
proximity.  We can see how the coronal temperatures and abundances of these
two key stars compare simply by plotting ratios of their coronal line
luminosities as a function of line formation temperature, as is done in
Figure~5.  The $\xi$~Boo~A line luminosities in the figure originate
from the line counts measured in Table~2.  The $\epsilon$~Eri
luminosities are computed from line counts similarly listed in WL06.

     Figure~5 shows that nearly all the line ratios are above one at
all temperatures, indicative of $\xi$~Boo~A being the more active star.
The ratios are clearly temperature dependent, with the ratio rising
from $\sim 1.5$ at $\log T=6.0$ to $\sim 5$ at $\log T=7.0$.
This implies a hotter corona for the more active $\xi$~Boo~A.  This is
consistent with previous observations demonstrating correlations between
coronal activity and coronal temperature \citep{mg97,mg04,at05}.
It is intriguing how smoothly and linearly the line ratios rise with
temperature.

     The lines in Figure~5 are divided into those with low first
ionization potential (FIP) and those with high FIP.  In the solar
corona and solar wind, elemental abundances are found to be dependent
on FIP.  Relative to the photosphere, elements with low FIP (Fe, Mg,
Si, etc.) are generally found to have coronal abundances that are
enhanced relative to elements with high FIP (C, N, O, Ne, etc.)
\citep{uf00}.  Evidence for this effect has been found for
some stars of low to moderate activity \citep{jml96,jjd97,jml99}.
However, on very active stars the
FIP effect tends to be either absent, or sometimes an inverse FIP effect
is observed, where low-FIP elements have coronal abundances that are
{\em lower} than photospheric
\citep{ma01,ma03,acb01,mg01,dph01,dph03,jsf03,jsf09,bb05}.

     Thus, for main sequence stars the picture is of the FIP
effect generally decreasing as activity increases \citep{at05}.
But there are exceptions to this trend, and Figure~5 is itself
inconsistent with a tight correlation between activity and
FIP effect, as the lower flux ratios of the high-FIP ions in Figure~5
demonstrate that $\xi$~Boo~A's corona exhibits a stronger FIP effect
than the less active $\epsilon$~Eri.  We will return to this issue in
\S4.6.

\subsection{Emission Measure Analysis}

     In order to quantify the coronal abundances and temperature
distributions, we perform an emission measure analysis for $\xi$~Boo~A
and B based on our measurements of their coronal emission lines.  The
EM analysis uses version 2.6 of the PINTofALE software developed by
\citet{vk98,vk00}.  This analysis follows closely that of
WL06 in our study of LETGS spectra from $\epsilon$~Eri, 36~Oph, and
70~Oph, so we do not describe it again in great detail here.

     Figure~6 shows EM distributions (in units of cm$^{-3}$) for the
two $\xi$~Boo stars, and compares them with distributions that we
previously computed for five other moderately active G8-K5 dwarfs
(WL06).  These distributions are computed using CHIANTI emissivities
\citep{kpd97,kpd09}, and ionization balance calculations from
\citet{pm98}.  Our line measurements are corrected for
interstellar absorption, but the interstellar H column densities
towards these stars, including $\xi$~Boo (see Table~1), are low enough
that this is not an important correction.

     For elements with detected lines we can compute coronal
abundances in the EM analysis.  However, line measurements alone only
allow {\em relative} abundances to be computed, and only the shape of
the EM distribution can be inferred as opposed to its absolute value.
Initially, we assume a solar photospheric abundance for Fe from
\citet{ng98}, and all other abundances are computed
relative to Fe.  The derived $\xi$~Boo abundance ratios are listed in
Table~3.

     In order to compute an {\em absolute} Fe abundance, and to
properly normalize the EM distribution, the line-to-continuum ratio
must be assessed.  Figure~7 illustrates this process for the case of
$\xi$~Boo~A.  The figure shows a synthetic spectrum generated from the
emission measure distribution in Figure~6, assuming an absolute Fe
abundance of ${\rm [Fe/H]}=0.7{\rm [Fe/H]}_{\odot}$, which is judged
to lead to the best fit.  This is the same as the ratio found for
$\epsilon$~Eri (WL06).  For $\xi$~Boo~B, we find a similar value of
${\rm [Fe/H]}=0.8{\rm [Fe/H]}_{\odot}$ (see Table~3).  These results
are used to properly normalize the EM distributions in Figure~6.
Figure~3 displays synthetic spectra of $\xi$~Boo~A and B based on the
EM distributions.  In addition to the dominant first-order spectrum,
we also take into account orders 2--5 in generating synthetic spectra.
Both Figures~3 and 7 explicitly show the contributions of these higher
orders to the total spectrum.

     In logarithmic terms, our measurement for $\xi$~Boo~A implies
${\rm \log [Fe/Fe_{\odot}]}=-0.15$, where ${\rm Fe_{\odot}}$ is here
the solar photospheric abundance of Fe.  This measurement can be
compared with the ${\rm \log [Fe/Fe_{\odot}]}=0.1^{+0.2}_{-0.15}$
measurement of \citet{jjd01} based on EUVE spectra.  At this point it
is worth noting that the contributions of weak, unresolved, and
unidentified lines to the observed ``continuum'' could easily lead to
underestimates of ${\rm [Fe/H]}$, so our ${\rm [Fe/H]}$ measurements
could be conservatively regarded as lower limits.  Given that our
measurement is lower than that of \citet{jjd01}, perhaps there are
more lines missing and unaccounted for within the LETGS spectral
region than in the EUVE spectral region analyzed by \citet{jjd01},
presumably leading to an underestimate of ${\rm [Fe/H]}$ on our part.
Nevertheless, both our measurement and that of \citet{jjd01} show a
higher Fe abundance in the corona than in the photosphere, where ${\rm
\log [Fe/Fe_{\odot}]}=-0.26$ (see Table~1), consistent with a
solar-like FIP effect (see \S4.3).  In following sections that focus
on comparing the coronal abundances of different stars, we avoid the
systematic errors involved in computing absolute abundances by
comparing only relative abundances, specifically abundances relative
to Fe.

     The seven EM distributions in Figure~6 are superficially similar
in appearance, with a dramatic rise at about $\log T=6.0$.  But
perhaps the most interesting feature is that all of the distributions
show an EM peak near $\log T=6.6$, a peak that is particularly convincing
for the two brightest stars, $\xi$~Boo~A and $\epsilon$~Eri.  The
coronal heating on these moderately active stars seems to favor this
particular temperature.

     The EM distribution that \citet{jml99} derived for
$\xi$~Boo~A from EUVE data agrees nicely with our {\em Chandra}
distribution.  Their peak at $\log T=6.6$ is not quite as narrow or
prominent, but this may just be a result of the ion-by-ion method that
\citet{jml99} use to quantify their emission measures.  As
there is some overlap in the EUVE and {\em Chandra}/LETGS spectra in
the $80-175$~\AA\ range, there are some lines in common to these two
analyses.  The most precise way to compare the activity level of
$\xi$~Boo~A during the EUVE and LETGS observations is to directly
compare the line fluxes rather than the derived EM distributions.  The
LETGS/EUVE line flux ratios have a weighted mean and standard deviation of
$0.75\pm 0.28$.

     An assessment of the relative flux calibration of EUVE and LETGS
by \citet{kb06} using white dwarf spectra suggests
calibration errors in the opposite direction, with LETGS/EUVE ratios
of $1.15\pm 0.07$.  Another systematic error that works in the
opposite direction is the fact that, unlike LETGS, the EUVE fluxes will
be contaminated by emission from $\xi$~Boo~B.  Thus, the below unity
value of the average LETGS/EUVE line ratio for $\xi$~Boo~A is likely
indicative of genuine stellar variability, with the star being
somewhat fainter during the LETGS observation.  However, the very
modest level of variability suggested by the LETGS/EUVE comparison is
hardly surprising, as the coronal emissions of cool main sequence
stars are known to be quite variable on all kinds of timescales.

     Finally, we can compute broadband X-ray luminosities directly
from the LETGS spectra.  We first subtract the higher order
contributions to the spectra estimated from the EM analysis (i.e., the
green line in Figure~7).  We can then combine the corrected spectra
with the known first order effective area curve and exposure time to
convert counts in our spectra to fluxes.  From the flux spectra we can
compute X-ray luminosities for any wavelength region within the
spectrum.  In Table~1, we list luminosities computed in this way for
the ROSAT-like $0.1-2.4$~keV ($5-120$~\AA) bandpass.  The total
$\xi$~Boo system luminosity of $\log L_X=28.98$ computed in this way
agrees reasonably well with the ROSAT all-sky survey measurement of
$\log L_X=28.91$ \citep{js04}.  The first order spectra suggest
$\xi$~Boo~A accounts for about 84.6\% of $\xi$~Boo's total X-ray flux,
compared to the 88.5\% contribution to counts in the zeroth order
image.  The discrepancy is due in part to the difference between count
measurements and flux, the latter of which requires consideration of
the effective area of the instrument.  The discrepancy is also due in
part to differences in the effective areas for the zeroth order image
and first order spectra.

\subsection{Coronal Abundances}

     In \S4.4, we compared the EM distributions of seven moderately
active G8-K5 dwarfs observed with LETGS.  We also wish to compare the
coronal abundances of these seven stars, in order to first determine
if a FIP effect exists for these stars, and if so to then compare its
magnitude within this stellar sample.  Our calculation of abundances
in the EM analysis assumes that abundances are the same at all
temperatures throughout the corona.  Figure~5 suggests that this is
probably a good assumption, as it shows that the differential FIP
effect between $\xi$~Boo~A and $\epsilon$~Eri is the same throughout
the $\log T=6.1-6.9$ temperature range that dominates the coronal
emission measure distributions of these stars.

     Figure~8 compares the coronal abundances of our
sample of stars, plotting the abundances relative to Fe, as listed
in Table~3.  But the search for a FIP effect relies on knowing not
just the coronal abundance by itself, but the coronal abundance
relative to the photospheric abundance.  Thus, in Figure~8 we subtract
photospheric abundances from the coronal ones.  However, because
photospheric abundances are imperfectly known, taking them into
account involves numerous assumptions, which we now describe.

     In Table~1, photospheric abundances measured for $\xi$~Boo~A are
listed for C, O, Mg, Si, and Fe.  These abundances are quoted relative
to solar.  The \citet{cap04} analysis from which these
abundances are obtained involves a line-by-line comparison of
solar and stellar photospheric absorption lines, so these stellar
abundances are {\em fundamentally} relative values.  For the purposes
of Figure~8, we require absolute photospheric abundances relative to
Fe, log~[X/Fe]$_*$, so we have to combine the abundances in Table~1 with
some assumed solar abundances.  In Figure~8a, we use solar abundances
from \citet{ng98}.

     There are two elements for which no stellar photospheric
abundances are available for $\xi$~Boo~A:  N and Ne.  Since these are
high-FIP elements, we assume that their abundances relative to the Sun
are the same as that of another high-FIP element, O.  In other words,
we assume ${\rm \log [N/N_{\odot}] = \log [Ne/Ne_{\odot}] = \log
[O/O_{\odot}]=-0.09}$ (see Table~1).  For $\xi$~Boo~B there are no
independent photospheric abundance measurements, so we simply have to
assume $\xi$~Boo~B is identical to $\xi$~Boo~A in this respect.
Similar assumptions have been made for the photospheric abundances of
$\epsilon$~Eri, 70~Oph~AB, and 36~Oph~AB (see WL06).

     The low-FIP and high-FIP data points in Figure~8a are
connected with separate dotted lines.  These dotted lines indicate
the following curious abundance patterns for all the
stars:  1. Mg abundances are always higher than those of Si, 2. Ne
abundances are always higher than the other high-FIP abundances, and
3. C and N abundances are intermediate between those of O and Ne.
Are these abundance patterns telling us something about how coronal
fractionation is operating in these moderately active stars?  Or are
these patterns just a product of assumptions made about the
photospheric abundances?

     At least for the high-FIP elements, the latter of these two
explanations seems most likely.  In Figure~8b, we explore this further
by changing our assumed photospheric abundances.  This figure is like
Figure~8a, but instead of \citet{ng98} solar abundances,
we use those of \citet{ma09}, with the exception of Ne.

     Photospheric Ne abundances provide unique problems, because there
are no photospheric Ne absorption lines to measure, which means that
solar Ne abundance measurements actually have to rely on emission
lines from the corona, where elements are subject to fractionation
processes like those that yield the FIP effect.  In Figure~8b, we
change the assumed photospheric abundance of Ne drastically by
assuming that the average coronal Ne/O ratio of ${\rm Ne/O}=0.41$
measured by \citet{jjd05} {\em for active stars} actually represents
the real {\em photospheric} abundance of {\em all} cool main sequence
stars, as opposed to the much lower solar coronal Ne/O ratio of ${\rm
Ne/O}=0.17\pm 0.05$ \citep{pry05,jts05}.  In other
words, rather than assuming that Ne is fractionated from O in active
stars but not in the Sun, in Figure~8b we assume that Ne is being
fractionated in the atmospheres of inactive stars like the Sun, but
not in active stars.  Theoretical support for this interpretation is
provided by \citet{jml09}, and a higher solar photospheric Ne abundance
could also help resolve discrepancies between solar interior models and
helioseismological observations \citep{hma05,jnb05}.

     With the revised assumptions regarding photospheric abundances
discussed above, the high-FIP abundance patterns seen in Figure~8a are
largely absent in Figure~8b, and the high-FIP dotted lines are
closer to horizontal.  This result could be used as evidence in favor
of the photospheric abundance assumptions being made in Figure~8b, as
opposed to the \citet{ng98} solar abundances being
assumed to make Figure~8a.

     Although changing the assumed solar abundances seems to be able
to significantly reduce the discrepancies among the high-FIP elements,
such is not the case for the low-FIP elements, where there is little
difference in the discrepancy between Mg and Si in Figures~8a and 8b.
Thus, presumably coronal fractionation processes really are enhancing
stellar Mg abundances relative to Si in the coronae of these
moderately active stars.  \citet{jml04,jml09} propose that coronal
fractionation originates from ponderomotive forces induced by the
propagation of Alfv\'{e}n waves through the chromosphere.  These
models demonstrate the potential for fractionating Mg relative to Si
in a manner consistent with the observed Mg/Si ratios.  In their
sample of early G dwarfs, \citet{at05} also generally find
high Mg/Si ratios.  But in a sample of M dwarfs, all of which happen
to show inverse FIP effects, \citet{cl08} find {\em low}
Mg/Si ratios, suggesting this behavior may be spectral type dependent.

\subsection{FIP Effect Variations}

     For the high-FIP dotted lines in Figure~8, the lower the line,
the higher the coronal Fe abundance is relative to the high-FIP
elements, and therefore the stronger the inferred FIP effect.
Significant FIP effect is apparent for many of the stars, with
70~Oph~A's corona possessing the strongest FIP bias.  But there is a
significant amount of scatter in the level of FIP effect found for this
sample of stars.  Figure~8 suggests the following sequence of
decreasing FIP bias: 70~Oph~A, $\xi$~Boo~A, 36~Oph~B, 36~Oph~A,
$\xi$~Boo~B, $\epsilon$~Eri, 70~Oph~B.  The last of these stars,
70~Oph~B, appears to actually have a slight {\em inverse} FIP effect.

     There is some spread of X-ray luminosity within our sample of
stars, but there is no correlation at all between X-ray luminosity and
FIP effect in this sample.  The most luminous star, $\xi$~Boo~A, shows
a strong FIP effect, but the second most luminous, $\epsilon$~Eri,
shows only a weak FIP effect at most.  The relatively strong FIP
effect of the most active star in our sample also runs contrary to the
general anti-correlation between stellar activity and FIP effect that
has been reported in the past (see \S4.3).

     Since the FIP effect on the Sun is latitude dependent, with
the effect being strong near the equator and minimal in the polar regions,
one might imagine that perhaps stellar orientation could
be a factor in the FIP bias observed from a star, assuming coronal
abundances on these more active stars also exhibit some sort of latitude
dependence.  However, since coronal emission is optically thin, at least
half of a star's total emission should be visible from a star regardless
of orientation, so it is hard to imagine orientation effects
explaining the roughly factor-of-3 variation in FIP bias seen in
Figure~8.

     In our previous paper (WL06) we have emphasized the remarkable
difference between 70~Oph~A and 70~Oph~B, which represent the FIP
effect extremes in Figure~8 despite being similar stars in the same
binary system.  One possible explanation that we proposed is that
perhaps the FIP effect is a time-dependent phenomenon.  The solar
example provides a precedent for this, with newly emerged active
regions having close to photospheric abundances and only acquiring a
FIP bias on timescales of days \citep{kgw01}.  Thus, for the stars in
Figure~8 with little or no apparent FIP bias (e.g., 70~Oph~B,
$\epsilon$~Eri), perhaps the visible surface of the star at the time
of observation was dominated by young active regions, whereas the
stars with strong observed FIP effects (e.g., 70~Oph~A, $\xi$~Boo~A)
were observed with older active regions.  The time-dependent
interpretation can only be tested by observing these stars again,
preferably more than once.  But the limited amount of evidence
available so far does {\em not} support this interpretation.
Abundances computed from EUVE spectra taken at different times from
our LETGS data seem to be very consistent with our LETGS results,
showing a strong FIP effect for $\xi$~Boo~A, and only a weak FIP
effect, if any, for $\epsilon$~Eri \citep{jml96,jml99}.  \citet{jsf03}
looked for abundance variations in multiple observations of the very
active K dwarf AB~Dor, with the observations encompassing a range of
activity levels, and found no such variability.

     There does not seem to be a strong case for activity dependence,
time dependence, or orientation effects being responsible for the
variation in FIP bias seen in Figure~8.  There does seem to be a
case for there being a spectral type dependence.  The two earliest
type stars ($\xi$~Boo~A and 70~Oph~A) have the strongest solar-like FIP
effect, while the three latest type stars (70~Oph~B, $\xi$~Boo~B, and
$\epsilon$~Eri) have the weakest.  This is explored further in
Figure~9, where we first compute the average of the high-FIP dotted
lines in Figure~8b, a quantity we simply call the ``FIP bias,'' and
then plot this quantity versus spectral type.

     In order to broaden the spectral range under consideration, in
Figure~9 we supplement our measurements with those of early G stars
from \citet{at05} and M dwarfs from \citet{cl08}.
In utilizing measurements from these other papers, we were careful to
compute the ``FIP bias'' in the same manner that we have for our
sample of stars, with the same assumptions about solar photospheric
abundances as in Figure~8b.  For the \citet{at05} sample,
stellar photospheric abundances are available from \citet{cap04},
which we take into account, but for the M
dwarfs of \citet{cl08} there are no such measurements and we
have no recourse but to assume that the photospheres of these stars
possess solar photospheric abundances.  Finally, we have added a point
for the Sun, using abundances listed in \citet{uf00}.

     Figure~9 shows a remarkably tight correlation of FIP bias with
spectral type, with the earlier type stars having solar-like FIP
effects and the later stars having inverse FIP effects, though the
lack of photospheric abundance measurements for the M dwarfs means
that conclusions about them are more uncertain.  The reversal
point seems to happen near spectral type K5~V.  To the best of our
knowledge, this is the first time a strong spectral type dependence of
FIP bias has been claimed.  Why has this seemingly strong correlation
not been recognized in the past?  Answering this question requires
clearly defining what the sample of stars in Figure~9 represents.  In
short, these are inactive to moderately active main sequence stars
with X-ray luminosities of $\log L_X\lesssim 29$.  For solar-like main
sequence stars such stars have rotational periods of $P_{rot}>6$~days
and ages of $t>0.35$~Gyr \citep{am87,jdd95,mg97,mg04}

     Most surveys of coronal X-ray emission are dominated by
interacting binaries, evolved stars, and other targets that represent
extremes of coronal activity, which are commonly observed by X-ray
telescopes because they are so bright and easy to detect.  But by
avoiding such extremes and focusing solely on more normal stars, we
find a very different empirical view of the FIP effect.  Gone is the
kind of activity dependence on which previous authors have focused
(see \S4.3), replaced instead with a strong spectral type dependence.
In order to emphasize the lack of activity dependence in Figure~9,
consider the two least active stars in the sample:  the Sun (G2~V) and
Proxima~Cen (M5.5~Ve).  The two G1 stars right next to the Sun in the
figure have X-ray luminosities almost two orders of magnitude higher
than the Sun.  But there is no difference in ``FIP bias''.  Likewise,
three of the four stars right next to Proxima~Cen in the figure have
X-ray luminosities about two orders of magnitude higher than it.  Once
again, no difference in FIP bias.

     There are two stars in the \citet{at05} sample that
are not included in Figure~9, because they have $\log L_X>29$ and
therefore violate our policy of avoiding extremes.  These two
stars are 47~Cas~B and EK~Dra, which both have rotation periods of
under three days.  This rapid rotation will probably exist for all
main sequence stars with $\log L_X>29$.  If we plotted points for
47~Cas~B and EK~Dra in Figure~9, they would lie well above the relation
defined by the other stars.  Rather than proposing that magnetic
activity is starting to operate somewhat differently at the activity
level of 47~Cas~B and EK~Dra, thereby yielding a different FIP bias,
we instead propose that the changes in FIP effect are ultimately
driven by changes to fundamental stellar properties induced by the
rapid stellar rotation.

     Rotation as rapid as that possessed by 47~Cas~B and EK~Dra will
presumably affect the atmospheric and convection zone properties of
the star to some extent \citep[e.g.,][]{bpb08}.  Interacting
binaries and evolved stars will also have substantial differences
between their atmospheric properties and those of main sequence stars
with identical spectral types.  Figure~9 suggests to us that FIP bias
is tied directly to basic stellar properties of the star, and not any
simple coronal property like total X-ray flux or magnetic field
strength.  Within the theoretical paradigm of \citet{jml04,jml09},
in which Alfv\'{e}n wave propagation through the chromosphere drives
coronal abundance anomalies, it is possible to imagine that changes in
basic stellar properties such as radius, surface gravity, or
convective motions could affect these waves and thereby affect the
coronal FIP bias.  {\em In short, we propose that the spectral type
dependence of FIP bias in Figure~9 provides a more fundamental insight
into the nature and cause of the FIP effect than the activity
trends that have been explored in the past.}

\section{SUMMARY}

     We have used {\em Chandra}/LETGS observations to study the
coronae of the $\xi$~Boo binary, and we have compared our results
with a previous study of five other moderately active GK dwarfs:
$\epsilon$~Eri, 70~Oph~AB, and 36~Oph~AB.  Our findings
are summarized as follows:
\begin{enumerate}
\item The two stellar components of $\xi$~Boo are resolved for the
  first time in X-rays.  We find that $\xi$~Boo~A accounts for 88.5\% of
  the counts in the zeroth order image.  For comparison, spectral
  analysis implies an 84.6\% contribution to the system's X-ray flux
  in the ROSAT-like 0.1-2.4~keV bandpass.
\item With $\xi$~Boo~B detected for the first time in X-rays, we now
  know it has an X-ray luminosity comparable to stars known to have
  strong stellar winds, leading to the conclusion that
  $\xi$~Boo~B may account for most of the modest wind observed from
  the system, with the more active $\xi$~Boo~A possessing a
  surprisingly weak wind despite its active corona.  This
  strengthens the case for the presence of a ``wind dividing line''
  separating moderately active stars with strong winds and very active
  stars with weak winds, possibly due to the increasing strength of
  large scale global magnetic fields that inhibit mass loss.
\item Our emission measure analysis of $\xi$~Boo~A and B finds evidence
  for a narrow peak in EM at $\log T=6.6$ for both stars, a peak also
  seen for the other moderately active G8-K5 dwarfs in our sample,
  indicating that the coronal heating mechanism operating on these
  stars favors this temperature.
\item The emission measure analysis provides us with coronal
  abundances relative to Fe.  Looking for coronal abundance anomalies
  requires correcting for stellar photospheric abundances, which in
  turn involves assuming solar photospheric abundances, since the
  stellar abundances we use are measured relative to solar.  When we
  simply assume solar abundances from \citet{ng98}, for
  our sample of stars we find that the high-FIP element abundances are
  inconsistent with each other.  This inconsistency is decreased
  significantly for all our stars if we instead assume a photospheric
  Ne abundance consistent with stellar coronal measurements from
  \citet{jjd05}, and solar abundances from \citet{ma09}
  for other elements.  This is an argument in favor of these latter
  photospheric abundances, unless there are actual coronal
  fractionation processes taking place among the high-FIP elements
  that we are erroneously removing.
\item Among low-FIP ions, the coronal Mg/Si ratio is
  consistently higher than the photospheric ratio for all of our
  G8-K5 stars, presumably an effect of actual coronal fractionation.  An
  enhancement of Mg relative to Si in the corona is predicted by many
  of the models of \citet{jml09}, which are designed to explain
  solar/stellar coronal abundance anomalies.  However, although
  \citet{at05} also finds high Mg/Si ratios for early G
  dwarfs, coronal Mg/Si ratios appear to be {\em lower} than
  photospheric for M dwarfs \citep{cl08}, suggesting a
  spectral type dependence for this coronal abundance ratio.
\item With regards to the FIP effect, our sample of stars exhibit
  a range of FIP bias, with 70~Oph~A and $\xi$~Boo~A having the
  strongest solar-like FIP effect, $\epsilon$~Eri showing little or no
  FIP effect, and 70~Oph~B possessing a slight inverse FIP effect.
  Expanding our sample of stars even further to include early G stars
  from \citet{at05} and M dwarfs from \citet{cl08},
  we find that the FIP bias correlates beautifully with spectral type,
  but only if extremely active stars with $\log L_X>29$ are excluded.
\item The tight correlation of FIP bias with spectral type suggests to
  us that the nature and causes of the FIP effect can ultimately be traced
  directly to fundamental surface properties.  Among normal main sequence
  stars, these change in a very predictable manner with spectral type,
  explaining the tight correlation in Figure~9.  No such correlation
  exists for extremely active stars, because many are evolved stars
  with disparate photospheric characteristics, and in other cases the
  stellar atmospheres will be affected by rapid rotation or the effects
  of close binarity.
\end{enumerate}

\acknowledgments

Support for this work was provided by the National Aeronautics and Space
Administration through Chandra Award Number GO8-9003Z issued by the Chandra
X-ray Observatory Center, which is operated by the Smithsonian
Astrophysical Observatory for and on behalf of the National Aeronautics
and Space Administration under contract NAS8-03060.

\clearpage

\begin{deluxetable}{lccc}
\tabletypesize{\scriptsize}
\tablecaption{Stellar Information}
\tablecolumns{4}
\tablewidth{0pt}
\tablehead{
  \colhead{Property\tablenotemark{a}} & \colhead{$\xi$~Boo~A} &
    \colhead{$\xi$~Boo~B} & \colhead{Refs.}}
\startdata
Other Name   & HD 131156 & HD 131156B & \\
Spect.\ Type &  G8 V     &   K4 V    & \\
Dist.\ (pc)  &  6.70     &   6.70    & 1\\
Radius (R$_{\odot}$)&0.83&   0.61    & 2\\
P$_{rot}$ (days)& 6.2   &   11.5    & 3\\
$\log L_{x}$\tablenotemark{b} & 28.86 & 27.97 & 4 \\
$\log L_{x}$\tablenotemark{c} & 28.91 & 28.17 & \\
$\dot{M}$ ($\dot{M}_{\odot}$)\tablenotemark{d} & 5 & 5 & 5 \\
$\log N_{H}$\tablenotemark{e} & 17.92 & 17.92  & 6 \\
$\log {\rm [C/C_{\odot}]}$  & $-0.10$ &   ...  & 7 \\
$\log {\rm [O/O_{\odot}]}$  & $-0.09$ &   ...  & 7 \\
$\log {\rm [Mg/Mg_{\odot}]}$ & $-0.26$ &   ... & 7 \\
$\log {\rm [Si/Si_{\odot}]}$ & $-0.10$ &   ... & 7 \\
$\log {\rm [Fe/Fe_{\odot}]}$ & $-0.26$ &   ... & 7 \\
\enddata
\tablenotetext{a}{The quantities in square brackets are stellar
  photospheric abundances relative to solar.}
\tablenotetext{b}{X-ray luminosities (ergs~s$^{-1}$) from ROSAT all-sky
  survey data, using the zeroth order LETGS image in Fig.~1 to establish
  the contributions of the individual stars.}
\tablenotetext{c}{X-ray luminosities in the ROSAT-like 0.1-2.4 keV
  (5-120~\AA) bandpass, computed directly from the LETGS spectra
  in Fig.~3.}
\tablenotetext{d}{Mass loss rate measurements from astrospheric
  absorption detections, where the measurement is for the
  combined mass loss from both stars of the binary.}
\tablenotetext{e}{Interstellar H~I column density.}
\tablerefs{(1) Perryman et al.\ 1997. (2) Barnes et al.\ 1978.
  (3) Noyes et al.\ 1984. (4) Schmitt \& Liefke 2004.
  (5) Wood et al.\ 2005a. (6) Wood et al.\ 2005b. (7) Allende Prieto
  et al.\ 2004.}
\end{deluxetable}

\clearpage

\begin{deluxetable}{lcccc}
\tabletypesize{\scriptsize}
\tablecaption{{\em Chandra} Line Measurements}
\tablecolumns{5}
\tablewidth{0pt}
\tablehead{
  \colhead{Ion} & \colhead{$\lambda_{rest}$} & \colhead{$\log T$} &
    \multicolumn{2}{c}{Counts} \\
  \colhead{} & \colhead{(\AA)} & \colhead{} &
    \colhead{$\xi$ Boo A} & \colhead{$\xi$ Boo B}}
\startdata
Si XIII  &   6.648 &6.99 & $  85.5\pm 21.0$ &          $<34.2$ \\
Si XIII  &   6.688 &6.99 &                  &                  \\
Si XIII  &   6.740 &6.99 &                  &                  \\
Mg XII   &   8.419 &7.11 & $  69.2\pm 17.2$ &          $<31.8$ \\
Mg XII   &   8.425 &7.11 &                  &                  \\
Mg XI    &   9.169 &6.80 & $ 207.7\pm 28.1$ &          $<33.2$ \\
Mg XI    &   9.231 &6.80 &                  &                  \\
Mg XI    &   9.314 &6.79 &                  &                  \\
Ne X     &  12.132 &6.87 & $ 362.2\pm 29.9$ & $  58.7\pm 15.8$ \\
Ne X     &  12.138 &6.87 &                  &                  \\
Fe XVII  &  12.264 &6.62 & $ 110.6\pm 21.8$ &          $<33.4$ \\
Fe XXI   &  12.285 &6.98 &                  &                  \\
Ne IX    &  13.447 &6.58 & $ 560.3\pm 38.6$ & $ 101.4\pm 21.9$ \\
Ne IX    &  13.553 &6.58 &                  &                  \\
Ne IX    &  13.699 &6.58 & $ 231.6\pm 28.7$ & $  77.0\pm 19.6$ \\
Fe XVIII &  14.203 &6.74 & $ 238.0\pm 30.3$ &          $<34.8$ \\
Fe XVIII &  14.208 &6.74 &                  &                  \\
Fe XVII  &  15.015 &6.59 & $1033.4\pm 43.8$ & $ 120.9\pm 20.7$ \\
Fe XVII  &  15.262 &6.59 & $ 514.9\pm 39.1$ & $  58.9\pm 18.6$ \\
O VIII   &  15.176 &6.65 &                  &                  \\
Fe XIX   &  15.198 &6.83 &                  &                  \\
O VIII   &  16.006 &6.63 & $ 465.5\pm 35.1$ & $  61.2\pm 18.3$ \\
Fe XVIII &  16.005 &6.73 &                  &                  \\
Fe XVIII &  16.072 &6.73 &                  &                  \\
Fe XVII  &  16.778 &6.58 & $ 598.2\pm 32.6$ & $  52.7\pm 15.1$ \\
Fe XVII  &  17.053 &6.58 & $1381.5\pm 46.2$ & $ 134.1\pm 20.8$ \\
Fe XVII  &  17.098 &6.58 &                  &                  \\
O VII    &  18.627 &6.34 & $  76.5\pm 17.4$ & $  20.5\pm 13.4$ \\
O VIII   &  18.967 &6.59 & $1227.3\pm 43.1$ & $ 215.2\pm 22.1$ \\
O VIII   &  18.973 &6.59 &                  &                  \\
O VII    &  21.602 &6.32 & $ 307.4\pm 25.5$ & $  76.9\pm 17.1$ \\
O VII    &  21.807 &6.32 & $  46.7\pm 16.1$ & $  14.5\pm 12.5$ \\
O VII    &  22.101 &6.31 & $ 191.9\pm 22.7$ & $  66.5\pm 18.1$ \\
N VII    &  24.779 &6.43 & $  83.8\pm 20.9$ &          $<32.0$ \\
N VII    &  24.785 &6.43 &                  &                  \\
C VI     &  33.734 &6.24 & $ 152.6\pm 20.9$ & $  41.6\pm 15.5$ \\
C VI     &  33.740 &6.24 &                  &                  \\
S XIII   &  35.667 &6.43 & $  45.3\pm 18.7$ &          $<30.0$ \\
Si XI    &  43.763 &6.25 & $  78.5\pm 19.0$ & $  14.0\pm 12.2$ \\
Si XII   &  44.019 &6.44 & $  87.5\pm 17.6$ & $  24.7\pm 13.0$ \\
Si XII   &  44.165 &6.44 & $ 183.3\pm 23.7$ & $  40.1\pm 16.8$ \\
Si XII   &  45.521 &6.44 & $  46.9\pm 18.1$ &          $<30.4$ \\
Si XII   &  45.691 &6.44 & $  96.3\pm 18.1$ &          $<31.4$ \\
Fe XVI   &  46.661 &6.43 & $  75.1\pm 20.2$ &          $<30.8$ \\
Fe XVI   &  46.718 &6.43 &                  &                  \\
Si XI    &  49.222 &6.24 & $  95.5\pm 20.2$ & $  23.9\pm 13.9$ \\
Fe XVI   &  50.361 &6.43 & $ 202.6\pm 24.0$ & $  13.3\pm 11.5$ \\
Fe XVI   &  50.565 &6.43 & $  80.5\pm 17.8$ &          $<27.4$ \\
Si X     &  50.524 &6.15 &                  &                  \\
Si X     &  50.691 &6.15 & $  16.8\pm 13.0$ &          $<25.8$ \\
Si XI    &  52.298 &6.24 & $  27.6\pm 13.6$ &          $<26.4$ \\
Fe XV    &  52.911 &6.32 & $  33.4\pm 14.3$ &          $<24.8$ \\
Fe XVI   &  54.127 &6.43 & $  48.2\pm 15.9$ &          $<27.2$ \\
Fe XVI   &  54.710 &6.43 & $ 115.4\pm 19.1$ &          $<25.4$ \\
Mg X     &  57.876 &6.22 & $ 101.0\pm 20.2$ &          $<29.4$ \\
Mg X     &  57.920 &6.22 &                  &                  \\
Fe XV    &  59.405 &6.32 & $  45.3\pm 15.1$ &          $<29.2$ \\
Fe XVI   &  62.872 &6.42 & $  54.5\pm 13.4$ &          $<26.0$ \\
Mg X     &  63.295 &6.21 & $  55.6\pm 13.4$ &          $<26.6$ \\
Fe XVI   &  63.711 &6.42 & $  94.1\pm 17.7$ & $  15.3\pm 10.6$ \\
Fe XVI   &  66.249 &6.42 & $ 188.6\pm 23.2$ &          $<27.4$ \\
Fe XVI   &  66.357 &6.42 &                  &                  \\
Fe XV    &  69.682 &6.32 & $ 137.0\pm 19.5$ & $  24.5\pm 12.5$ \\
Fe XV    &  69.941 &6.32 & $  59.1\pm 18.0$ &          $<29.4$ \\
Fe XV    &  69.987 &6.32 &                  &                  \\
Fe XV    &  70.054 &6.32 &                  &                  \\
Mg IX    &  72.312 &6.02 & $  30.0\pm 13.4$ & $  19.0\pm 11.7$ \\
Fe XV    &  73.472 &6.32 & $  74.4\pm 18.4$ & $  19.5\pm 11.8$ \\
Fe XVI   &  76.497 &6.42 & $  46.6\pm 14.6$ &          $<28.6$ \\
Ne VIII  &  88.082 &5.96 & $  65.5\pm 16.5$ & $  35.0\pm 15.8$ \\
Ne VIII  &  88.120 &5.96 &                  &                  \\
Fe XVIII &  93.923 &6.68 & $ 309.2\pm 29.5$ & $  25.0\pm 16.7$ \\
Ne VIII  &  98.116 &5.94 & $  45.6\pm 18.0$ &          $<40.2$ \\
Ne VIII  &  98.260 &5.94 & $  62.1\pm 21.0$ & $  27.2\pm 19.3$ \\
Fe XIX   & 101.550 &6.80 & $  35.5\pm 17.5$ &          $<39.0$ \\
Fe XVIII & 103.937 &6.68 & $  84.2\pm 21.6$ &          $<38.8$ \\
Fe XIX   & 108.355 &6.79 & $ 158.5\pm 25.5$ &          $<40.2$ \\
Fe XXII  & 117.180 &7.03 & $  75.0\pm 22.6$ &          $<39.6$ \\
Fe XX    & 121.845 &6.88 & $  60.7\pm 18.5$ &          $<39.8$ \\
Fe XXI   & 128.752 &6.95 & $  46.2\pm 18.0$ &          $<40.8$ \\
Fe XX    & 132.840 &6.88 & $ 138.8\pm 26.3$ &          $<39.2$ \\
Fe XXIII & 132.906 &7.12 &                  &                  \\
Fe IX    & 171.073 &5.95 & $  31.0\pm 16.4$ & $  27.2\pm 17.9$ \\
\enddata
\end{deluxetable}

\clearpage

\begin{deluxetable}{cccc}
\tabletypesize{\normalsize}
\tablecaption{Coronal Densities and Abundances}
\tablecolumns{4}
\tablewidth{0pt}
\tablehead{
  \colhead{} & \colhead{$\xi$ Boo A} & \colhead{$\xi$ Boo B} &
    \colhead{Sun\tablenotemark{a}}}
\startdata
$\log n_{e}$    & $<10.24$ & $<11.43$ & 9.4 \\
${\rm Fe/Fe_{\odot}}$ & 0.7 & 0.8 & 1 \\
$\log {\rm [Fe/H]}$  & $-4.65$ & $-4.60$ & $-4.50$ \\
$\log {\rm [C/Fe]}$  & $0.75_{-0.07}^{+0.09}$ & $0.87_{-0.34}^{+0.28}$ &
  1.02 \\
$\log {\rm [N/Fe]}$  & $0.19_{-0.17}^{+0.11}$ & ...                    &
  0.42 \\
$\log {\rm [O/Fe]}$  & $1.02_{-0.03}^{+0.03}$ & $1.13_{-0.12}^{+0.12}$ &
  1.33 \\
$\log {\rm [Ne/Fe]}$ & $0.58_{-0.03}^{+0.02}$ & $0.87_{-0.15}^{+0.09}$ &
  0.58 \\
$\log {\rm [Mg/Fe]}$ & $0.23_{-0.06}^{+0.08}$ & $0.22_{-0.98}^{+0.38}$ &
  0.08 \\
$\log {\rm [Si/Fe]}$ & $0.04_{-0.07}^{+0.07}$ & $0.20_{-0.36}^{+0.18}$ &
  0.05 \\
$\log {\rm [S/Fe]}$  &$-0.28_{-0.57}^{+0.24}$ & ...                    &
  $-0.17$\\
\enddata
\tablenotetext{a}{The coronal density quoted for the Sun is from
  McKenzie (1987), while the abundances listed are photospheric abundances
  from Grevesse \& Sauval (1998).}
\end{deluxetable}

\clearpage

\begin{figure}[t]
\plotfiddle{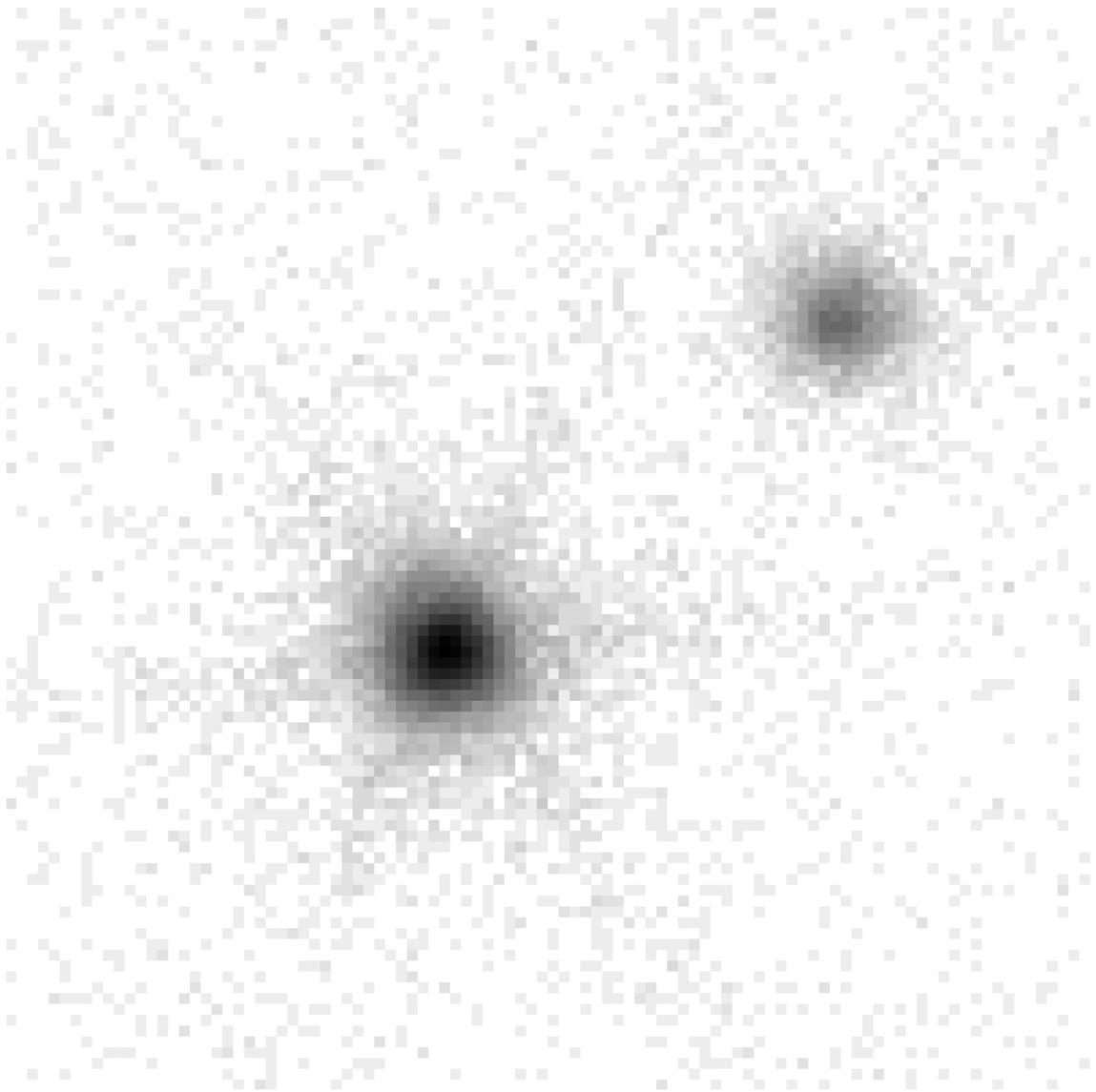}{5.0in}{0}{80}{80}{-315}{-30}
\caption{Zeroth-order image of the $\xi$~Boo binary (G8~V+K4~V) from a
  {\em Chandra} LETGS observation, with north being up in the figure.
  The brighter G8~V component of the binary accounts for 88.5\% of the
  system's total X-ray counts.  The stellar separation is
  $6.30^{\prime\prime}$.}
\end{figure}

\clearpage

\begin{figure}[t]
\plotfiddle{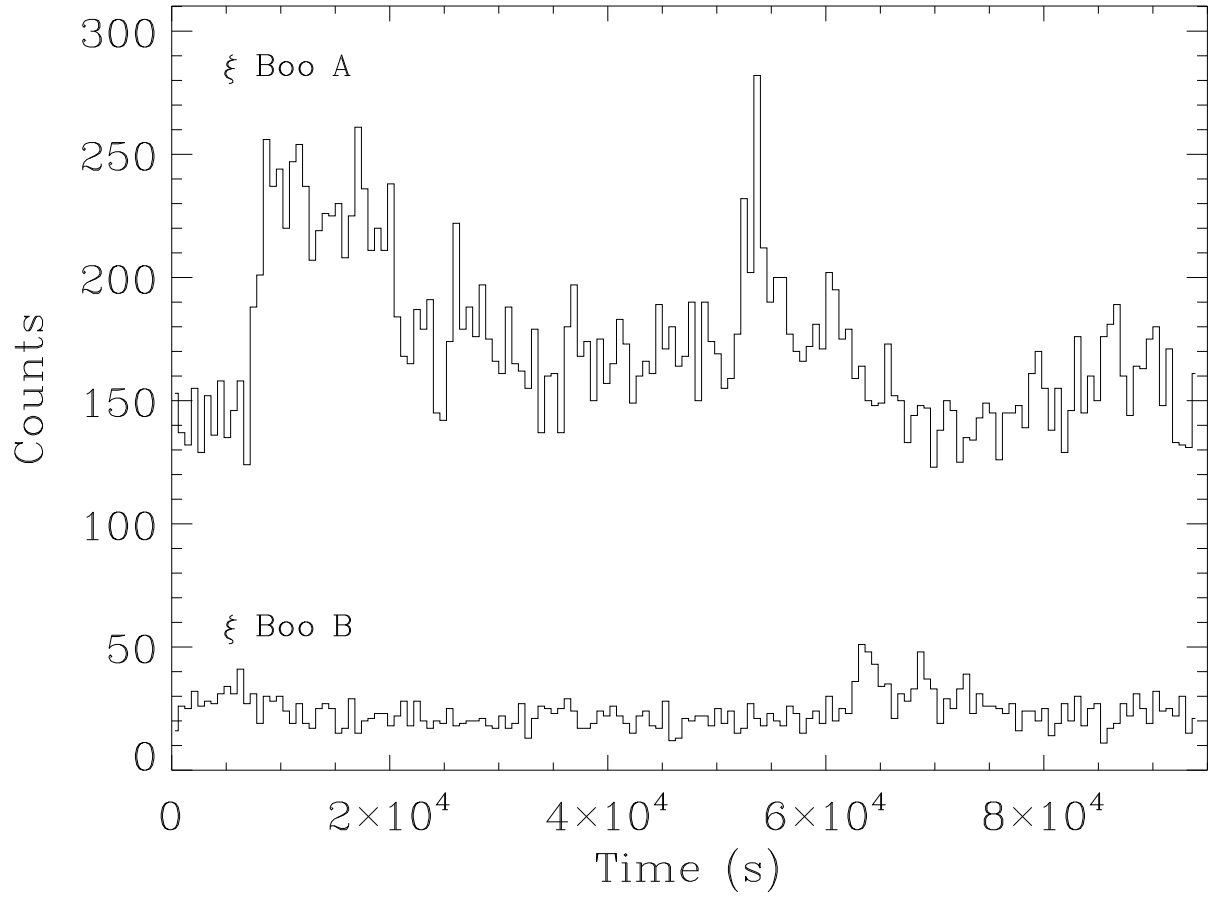}{4.5in}{90}{75}{75}{280}{-60}
\caption{X-ray light curves for $\xi$~Boo A and B
  measured from the zeroth-order images in the {\em Chandra} LETGS
  data, using 10 minute time bins.}
\end{figure}

\clearpage

\begin{figure}[t]
\plotfiddle{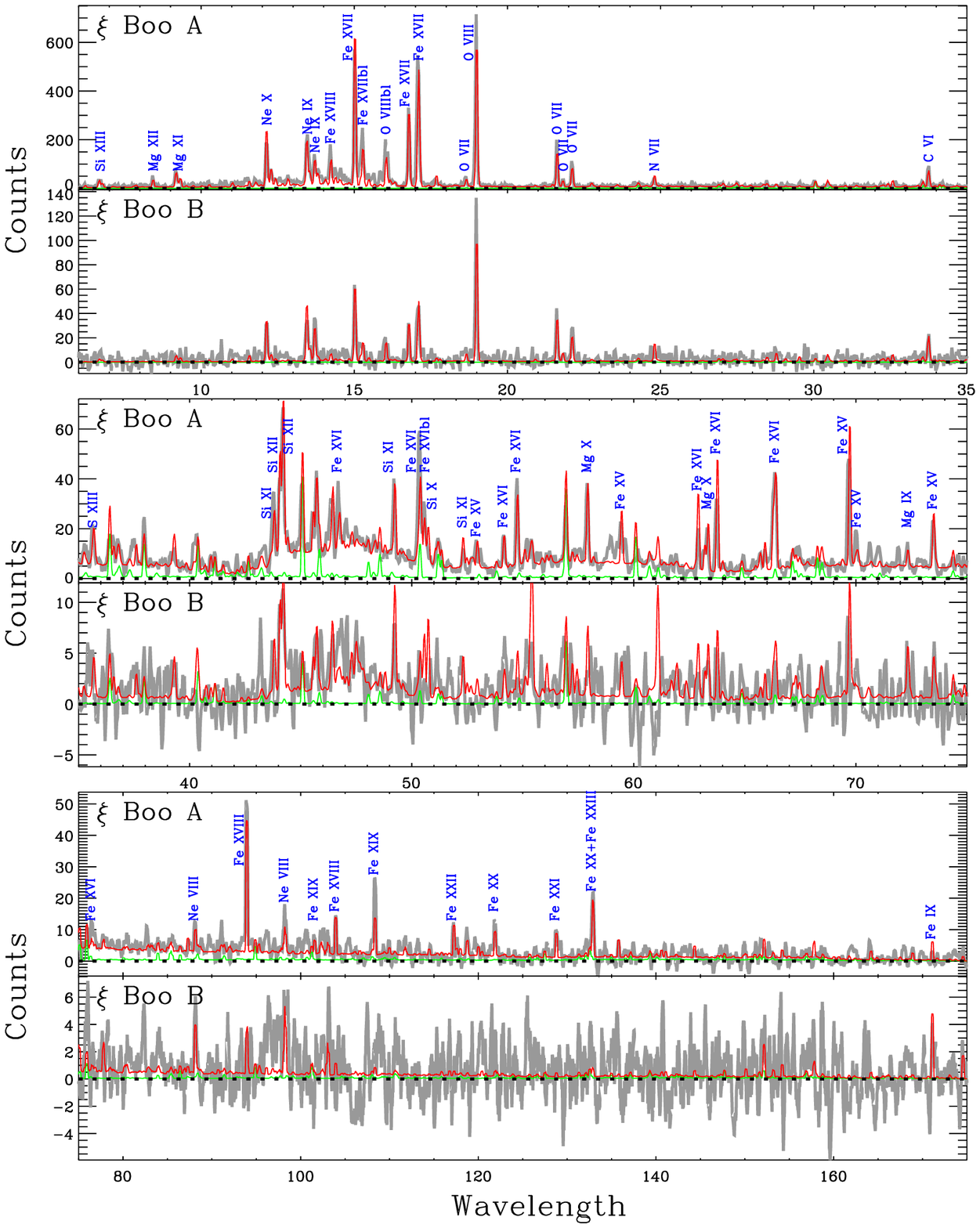}{7.0in}{0}{95}{95}{-310}{-110}
\caption{The {\em Chandra} LETGS spectra of
  $\xi$~Boo A and B, rebinned by a factor of 3 to improve S/N.  For
  wavelengths above 35~\AA, the spectra are also smoothed for the sake
  of appearance.  The red lines are synthetic spectra computed using
  derived emission measure distributions (see Fig.~6),
  and the green lines indicate the contributions of higher spectral orders
  (2--5) to the model spectra.}
\end{figure}

\clearpage





\begin{figure}[t]
\plotfiddle{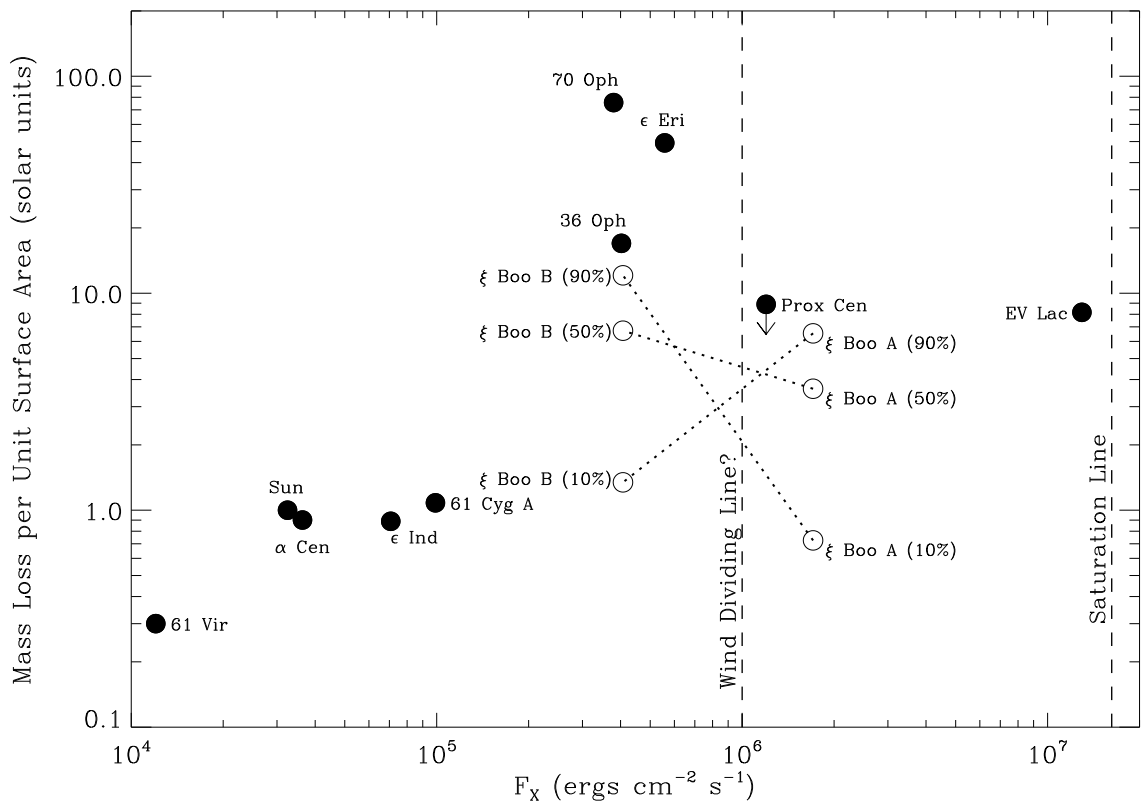}{2.5in}{0}{90}{90}{-280}{-340}
\caption{Mass loss rates (from Wood et al.\ 2005)
  plotted versus X-ray surface flux for various main sequence stars.
  Since only $\xi$~Boo's total mass loss rate is known, points are plotted
  for $\xi$~Boo~A and B assuming 3 different divisions of wind between
  the two stars.  Only if $\xi$~Boo~B accounts for most of the binary's
  wind is $\xi$~Boo~B consistent with similar stars (especially 36~Oph).
  In this instance $\xi$~Boo~A's wind would have to be very weak,
  strengthening the case for a ``wind dividing line'' marking
  where the relation between coronal winds and X-ray emission changes
  dramatically.}
\end{figure}

\clearpage

\begin{figure}[t]
\plotfiddle{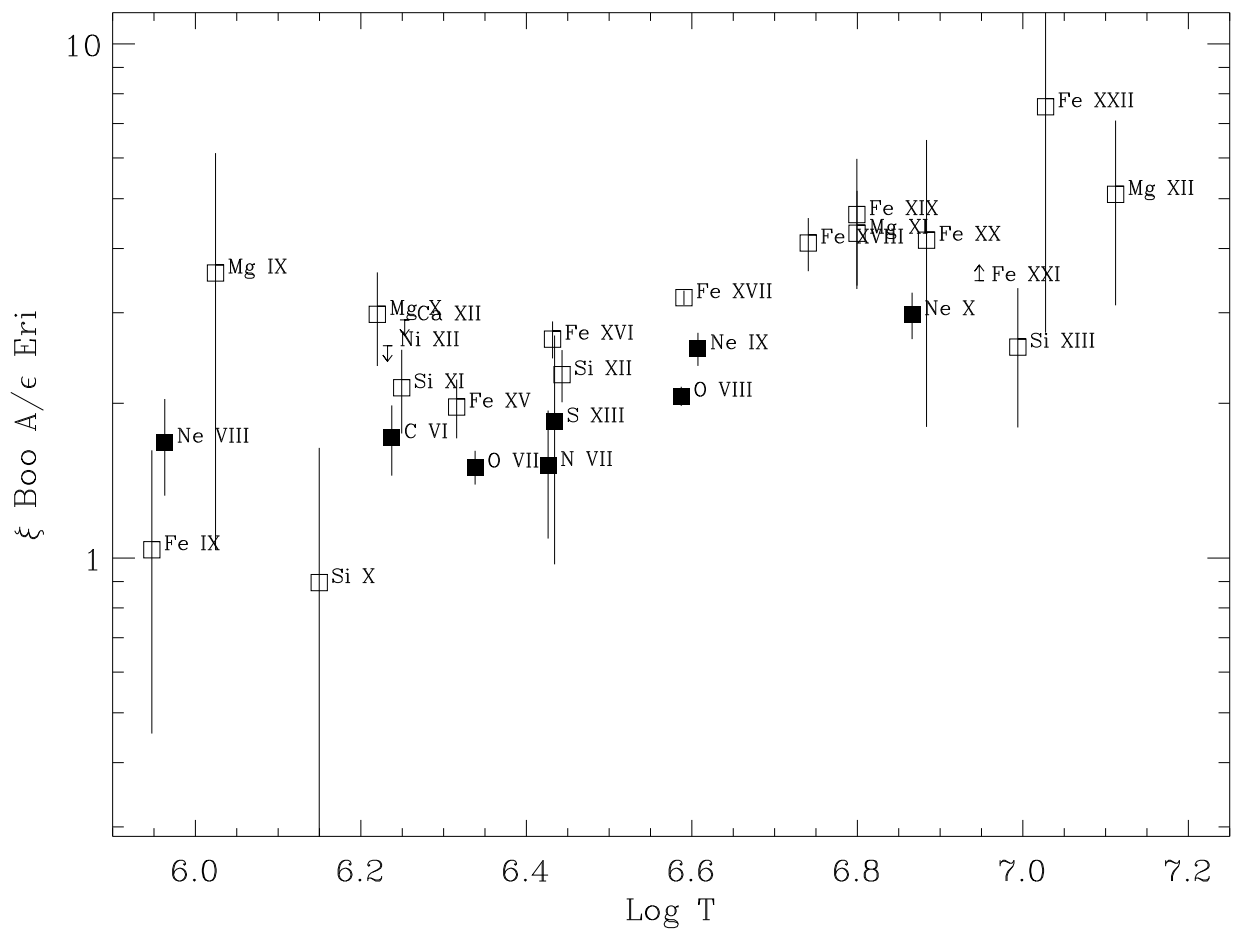}{3.0in}{90}{75}{75}{290}{-60}
\caption{$\xi$~Boo~A/$\epsilon$~Eri line luminosity ratios for
  emission lines detected for both stars, plotted versus line formation
  temperature.  The observed ratios above one indicate that $\xi$~Boo~A is
  somewhat more active than $\epsilon$~Eri at all temperatures.
  The ratios smoothly increase with temperature, implying a hotter
  corona for $\xi$~Boo~A.  Higher ratios are clearly observed
  for lines of low FIP elements (open boxes) than for high FIP
  elements (filled boxes), indicating that the corona of $\xi$~Boo~A
  has a stronger FIP effect than $\epsilon$~Eri.}
\end{figure}

\clearpage

\begin{figure}[t]
\plotfiddle{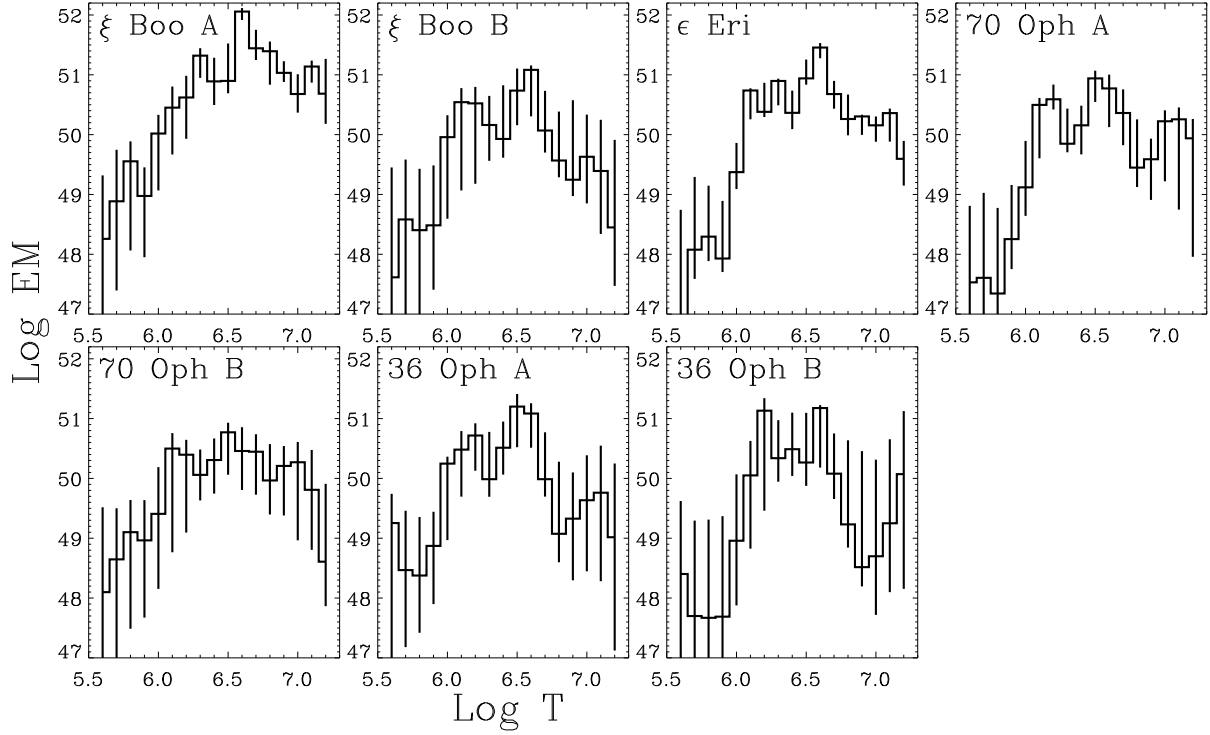}{3.3in}{90}{70}{70}{270}{-90}
\caption{Emission measure distributions derived for
  seven moderately active G8-K5 dwarfs observed by {\em Chandra} LETGS,
  with 90\% confidence error bars.  The $\xi$~Boo~AB analyses are
  from this paper, while the other five are from WL06.
  Note the peak near $\log T=6.6$
  seen for all these stars.}
\end{figure}

\clearpage

\begin{figure}[t]
\plotfiddle{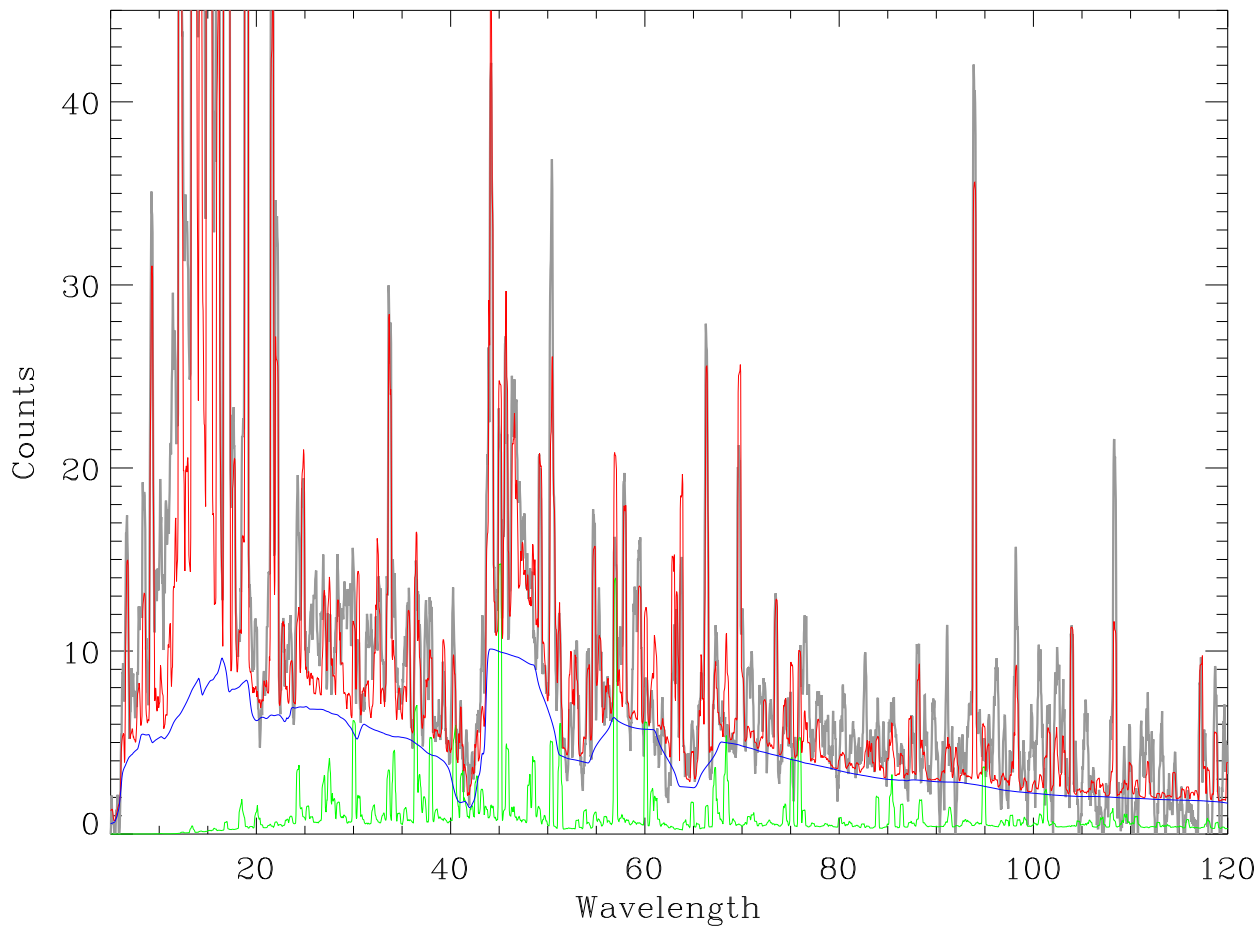}{3.3in}{90}{75}{75}{290}{-60}
\caption{A synthetic spectrum (red line) showing
  our best fit to the highly smoothed {\em Chandra} $\xi$~Boo~A spectrum
  (gray line).  Line contributions are computed from the emission measure
  distribution in Fig.~6.  The continuum contribution
  to the total spectrum (blue line) assumes an absolute Fe abundance of
  ${\rm [Fe/H]}=0.7{\rm [Fe/H]}_{\odot}$.  The green line shows the
  contributions of higher spectral orders (2-5) to the total spectrum.}
\end{figure}

\clearpage

\begin{figure}[t]
\plotfiddle{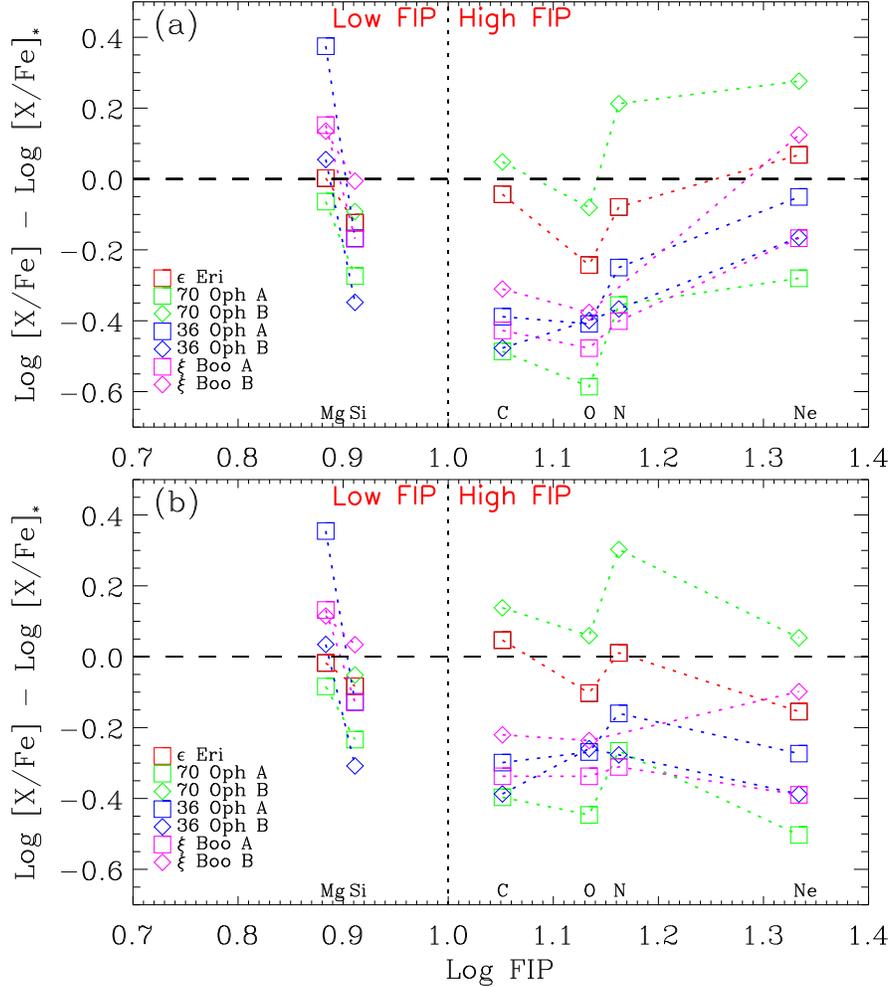}{4.0in}{0}{110}{110}{-325}{-425}
\caption{(a) Coronal abundances of various elements relative to Fe
  plotted versus FIP (in eV), for seven moderately active G8-K5 dwarf
  stars.  The abundance ratios are shown relative to assumed stellar
  photospheric ratios ($\log {\rm [X/Fe]_*}$).  The vertical dotted
  line separates low-FIP and high-FIP elements, and colored dotted
  lines connect the abundances for each star within these two regimes.
  The high-FIP abundance levels suggest the following sequence of
  increasing FIP effect: 70~Oph~B, $\epsilon$~Eri, $\xi$~Boo~B,
  36~Oph~A, 36~Oph~B, $\xi$~Boo~A, 70~Oph~A.  (b) Same as (a), but
  with different assumptions about the photospheric abundances.
  Instead of Grevesse \& Sauval (1998) as the source of solar
  abundances,we here use Drake \& Testa (2005) for Ne, and Asplund et
  al.\ (2009) for other elements.  With these new assumptions, the
  high-FIP dotted lines are more horizontal, and therefore more
  self-consistent, possibly indicating that the photospheric
  abundances assumed here are better than in (a).}
\end{figure}


\begin{figure}[t]
\plotfiddle{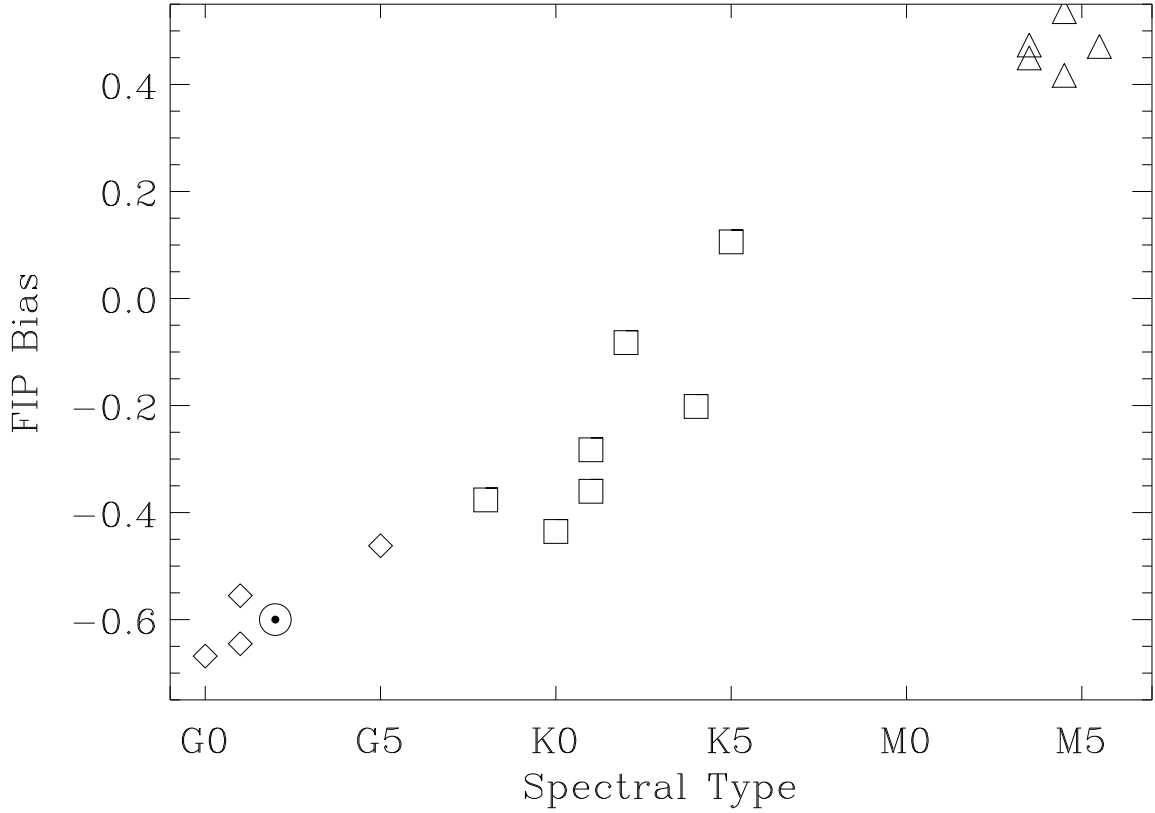}{3.0in}{0}{95}{95}{-300}{-350}
\caption{The boxes indicate the average values of the high-FIP curves
  from Fig.~8b, here simply called the ``FIP bias,'' plotted versus
  spectral type.  We supplement our measurements with ones from
  Telleschi et al.\ (2005) (diamonds) and Liefke et al.\ (2008)
  (triangles), in addition to a solar value from Feldman \& Laming
  (2000).  For all the GK stars the FIP bias calculations include
  corrections for stellar photospheric abundances from Allende Prieto
  et al.\ (2004), but for the M stars there are no stellar
  photospheric measurements available so we have to simply assume
  solar photospheric abundances apply.  The uniform assumption of
  abundances other than solar photospheric for the M dwarfs could
  potentially affect the consistency of the M dwarfs with the
  correlation defined by the other stars.  For purposes of this
  figure, we avoid extremes of stellar activity, confining our
  attention to stars with $\log L_X < 29$.}
\end{figure}

\end{document}